\documentstyle[epsfig,english,rotating,11pt,a4,cite]{article}
\renewcommand{\Huge}{\huge}
\def\ifmath#1{\relax\ifmmode #1\else $#1$\fi}%
\def\GeV{\ifmmode \hbox{\rm Ge\kern -0.1em V}\else
                  \hbox{\mathrm{Ge\kern -0.1em V}}\fi}%
\def\MeV{\ifmmode \hbox{\rm Me\kern -0.1em V}\else
                  \hbox{\mathrm{Me\kern -0.1em V}}\fi}%
\def\keV{\ifmmode \hbox{\rm ke\kern -0.1em V}\else
                  \hbox{\mathrm{ke\kern -0.1em V}}\fi}%
\def\eV{\ifmmode \hbox{\rm e\kern -0.1em V}\else
                 \hbox{\mathrm{e\kern -0.1em V}}\fi}%

\newcommand {\Alrfb}   {\tilde{\rm {A}}_{FB}}
\newcommand {\Alrfbb}   {\tilde{\rm {A}}_{FB}^{\rm b}}
\newcommand {\Afbpol}     {A^{0,\,\ell}_{\rm {FB}}}
\newcommand {\Afbzl}     {A^{0,\,\ell}_{\rm {FB}}}
\newcommand {\Afbze}     {A^{0,\,{\rm e}}_{\rm {FB}}}
\newcommand {\Afbzf}     {A^{0,\,{\rm f}}_{\rm {FB}}}
\newcommand {\Afbzm}     {A^{0,\,\mu}_{\rm {FB}}}
\newcommand {\Afbzt}     {A^{0,\,\tau}_{\rm {FB}}}
\newcommand {\Afbzb}     {A^{0,\,{\rm b}}_{\rm {FB}}}
\newcommand {\Afbzc}     {A^{0,\,{\rm c}}_{\rm {FB}}}

\newcommand {\sigmal}  {\sigma_L}
\newcommand {\sigmar}  {\sigma_R}

\renewcommand {\rm} {\mathrm}

\def\leqsim{\mathbin{\;\raise1pt\hbox{$<$}\kern-8pt\lower3pt\hbox{$\sim$}\;}}
\def\geqsim{\mathbin{\;\raise1pt\hbox{$>$}\kern-8pt\lower3pt\hbox{$\sim$}\;}}

\newcommand {\Ab}      {\rm{A_b}}
\newcommand {\Ac}      {\rm{A_c}}
\newcommand {\chibar}     {\bar{\chi}}

\newcommand {\bbbar}        {{\rm{b}\bar{\rm{b}}}}

\newcommand {\qqbar}        {{\rm{q}\bar{\rm{q}}}}

\newcommand {\Rqqz}       {{R^{0}_{\mathrm{q}}}}
\newcommand {\Rccz}       {{R^{0}_{\mathrm{c}}}}
\newcommand {\Rbbz}       {{R^{0}_{\mathrm{b}}}}
\newcommand {\chit}       {\chi^{2}}
\newcommand {\Ae}         {a_{\rm{e}}}
\newcommand {\Abcoup}     {a_{\rm{b}}}
\newcommand {\Accoup}     {a_{\rm{c}}}
\newcommand {\Af}         {a_{\rm{f}}}
\newcommand {\Alll}       {a_{\ell}}
\newcommand {\Vnu}        {v_{\nu}}

\newcommand {\Ve}         {v_{\rm{e}}}
\newcommand {\Vf}         {v_{\rm{f}}}
\newcommand {\Vlll}       {v_{\ell}}
\newcommand {\Vb}         {v_{\rm{b}}}
\newcommand {\Vc}         {v_{\rm{c}}}
\newcommand {\Aferm}      {A_{\rm f}}
\newcommand {\Aelec}      {A_{\rm e}}

\newcommand {\Aell}       {A_{\ell}}

\newcommand {\Gf}      {{\Gamma_{\rm f}}}
\newcommand {\Gb}      {{\Gamma_{\rm b}}}
\newcommand {\Gc}      {{\Gamma_{\rm c}}}

\newcommand {\Gl}      {{\Gamma_{l}}}
\newcommand {\Ge}      {{\Gamma_{\rm e}}}

\renewcommand{\thefootnote}{\fnsymbol{footnote}}

\newcommand {\alphamz}    {\alpha(\MZ)}
\newcommand {\alphasmz}    {\alpha_{\rm{s}}{\rm(\MZ)}}

\newcommand {\MZ}      {M_{\mathrm{Z}}}
\newcommand {\MW}      {M_{\mathrm{W}}}

\newcommand {\MH}      {m_{\mathrm{H}}}

\newcommand {\Mt}      {m_{\mathrm{t}}}

\newcommand {\GZ}      {\Gamma_{\mathrm{Z}}}

\newcommand {\Ree}      {R_{\mathrm{e}}}
\newcommand {\Rmu}      {R_{\mu}}
\newcommand {\Rtau}      {R_{\tau}}
\newcommand {\Rl}      {R_{\ell}}

\newcommand {\thw}        {\theta_{\mathrm{W}}}

\newcommand {\swsqa}       {\sin^2\!\thw}

\newcommand {\swsqeffff}   {\sin^2\!\theta_{\rm{eff}}^{\rm {f}}}
\newcommand {\swsqeffl}    {\sin^2\!\theta_{\rm{eff}}^{\rm {lept}}}

\newcommand {\ff}         {{\rm f}\overline{\rm f}}

\newcommand {\bb}    {{\mathrm b\overline{\mathrm b}}}
\newcommand {\cc}    {{\rm c\overline{\rm c}}}

\newcommand {\Ghad}       {\Gamma_{\mathrm{had}}}

\def\shad{\sigma_{\mathrm{h}}^{0}}

\newcommand {\ALR} {\mbox{$A_{\rm {LR}}$}}

\newcommand {\cAe} {\mbox{$\cal A_{\rm e}$}}
\newcommand {\cAt} {\mbox{$\cal A_{\tau}$}}

\newcommand {\cAl} {\mbox{$\cal A_{\ell}$}}
\newcommand {\cAb} {\mbox{$\cal A_{\rm b}$}}
\newcommand {\cAc} {\mbox{$\cal A_{\rm c}$}}

\newcommand {\Abb}   {\ifmath{A_{\mathrm{FB}}^{\mathrm{b\bar{b}}}}}

\newcommand {\Itf}         {I^3_{\rm f}}
\newcommand {\Qf}          {Q_{\rm f}}

\newcommand {\rhof}        {\rho_{\rm f}}

\def\Ups4s{\mbox{$\Upsilon(4S)$}}
\newcommand {\HI} {\mbox{$90 \le \MH~[{\rm {GeV}}] \le 1000$}}
\newcommand {\tI} {\mbox{$168.8 \le \Mt~[{\rm {GeV}}] \le 178.8$}}

\newcommand{\avQfb} {\langle{\rm Q}_{\rm FB}\rangle}
\input{rotate}
\begin{document}
\flushbottom
\begin{titlepage}
\begin{flushright}
       OUNP-98-08 \\
       hep-ph/9811415 \\
\end{flushright}
\begin{center}
\boldmath
\Huge {\bf Are there anomalous Z fermion couplings ?
}\\
\unboldmath
\vspace*{0.8cm}
\Large {\bf Peter B. Renton} \\
\vspace*{0.6cm}
\Large{Particle and Nuclear Physics Laboratory} \newline
\Large{University of Oxford, Oxford OX1 3RH, U.K.} \\
\Large{ e-mail: p.renton@physics.ox.ac.uk}
\end{center}

\vspace*{0.4cm}
\begin{abstract}
 The couplings of the fermions to the Z boson are of great importance in
establishing the validity of the Standard Model and in looking for
physics beyond it.

The couplings of the b-quark to the Z boson have been the subject of much
experimental study and theoretical interpretation. The apparent 
excess in
the value of $\Rbbz$, the ratio of the partial width of the Z boson 
to $\bbbar$ to its total hadronic width, above the Standard Model expectation
reported a few years ago has now become much less significant. 
However, the measurements of the
pole forward-backward asymmetry $\Afbzb$ for b-quarks at the Z pole
and of the {\it polarisation
parameter} $\Ab$, obtained using a polarised electron beam, have
improved considerably in accuracy. 

 The latest data are examined and values of the vector and axial-vector
b-quark and c-quark couplings to the Z are extracted. The left and right handed
couplings are also extracted. It is found that whereas the c-quark couplings 
are compatible with the Standard Model, those of the b-quark data are only
compatible with the Standard Model at about the 1\% level. In addition, the
individual lepton couplings are extracted and the hypothesis of {\it
lepton universality} is examined.

 The sensitivity of the limits from electroweak fits to the Higgs boson mass 
to these data is examined.
 
\end{abstract}

\end{titlepage}
\renewcommand{\thefootnote}{\arabic{footnote}}
\setcounter{footnote}{0}


\section{Introduction}

 The couplings of leptons and quarks to the Z boson are of fundamental 
importance
both in testing the Standard Model (SM) and in searching for, or setting
limits on, physics beyond the SM.

 The results available in the Summer of 1995 \cite{beijing} showed some
possibly significant differences from the SM expectations in the
values of $\Rbbz$ and $\Rccz$; where $\Rqqz$ is the ratio of the Z partial
width to $\qqbar$ to the total hadronic width. These were about 3.1 
and 2.4 standard deviations above and below the SM values for $\Rbbz$ 
and $\Rccz$ respectively. 

The Z$\bbbar$ vertex is a sensitive probe for new physics arising from
vertex corrections.
The above results, in particular those for $\Rbbz$, gave rise to considerable
theoretical speculation on possible physics beyond the SM which could
lead to such an increase. Within the context of Supersymmetry a possible
explanation was the existence of light Supersymmetric particles (charginos, 
top-squarks), which could suitably enhance $\Rbbz$. 
However, the apparently low value for $\Rccz$ did not
seem to have a straightforward interpretation.

 Since that time the data samples of Z bosons analysed, both at the 
CERN LEP accelerator and at the SLC, have considerably increased. 
The experimental techniques
used in b and c-tagging have also improved significantly due to the advent
of improved Microvertex Detectors and the implementation of new tags.
The  measurements of the
forward-backward asymmetry $\Afbzb$ for b-quarks at the Z pole
and of the {\it polarisation
parameter} $\Ab$, obtained using a polarised electron beam, have
improved considerably in accuracy. 

 The latest, but still largely preliminary, data are examined and values 
of the vector and axial-vector
b-quark and c-quark Z couplings are extracted. The left and right handed
couplings are also extracted. These couplings are compared to those
expected in the Standard Model.

 The hypothesis of {\it lepton universality} is built into the SM. 
In the fits to the current data set discussed below, the degree to which the 
data support this hypothesis is discussed.

 One important aspect of precision electroweak fits is on the
constraints the data give on the Standard Model Higgs boson.
The sensitivity of the limits from electroweak fits to the Higgs boson mass 
to the most sensitive data is examined. In these fits the precise value
of the top-quark mass $\Mt=173.8\pm5.0$~GeV measured at Fermilab by the CDF and
D0 experiments\cite{topmass} is a very important constraint.

\section{Z boson couplings}

 The couplings of a fermion f to the Z boson are specified by its
effective vector and axial-vector couplings $\Vf$ and $\Af$ respectively.
A useful quantity is the
{\it coupling} or {\it polarisation parameter} for fermion f
\begin{equation}
 A_{f} = \frac{2\Vf\Af}{\Vf^2 + \Af^2} .
\end{equation}

The pole forward-backward asymmety of fermion f is given by
\begin{equation}
 \Afbzf = \frac{3}{4}A_{e}A_{f} .
\end{equation}

 Since $\Aferm$
depends on the ratio $\Vf$/$\Af$, a measurement of $\Afbzf$ depends on
both $\Ve/\Ae$ and $\Vf/\Af$. The effective couplings can also be written as
\begin{equation}
     \Af = \Itf \sqrt{\rhof} \ , \ \ \ \ \
 \frac{\Vf}{\Af} = 1 - 4|\Qf| \swsqeffff \ ,
\end{equation}
where the mixing angle defined for {\it leptons} ($\swsqeffl$) is
used for reference.
Those defined for quarks have  small shifts, due to SM plus any
new physics~\cite{ORR}.
The Z partial decay width to $\ff$ is $\Gf$ $\sim (\Vf^2 + \Af^2 )$.
\footnote{In addition there is a much smaller term proportional
to ($\Vf^2 - \Af^2$)m$_{f}^{2}/\MZ^2$. This is taken into account in
the fits discussed here.} 

 The results of measurements of the $\tau$ polarisation are very sensitive to
the lepton couplings. The average $\tau$ polarisation gives A$_{\tau}$ and
the forward-backward asymmetry gives  A$_{e}$.

 The results of the hadronic and leptonic cross-sections and the leptonic
forward-backward asymmetries are expressed in terms of the five 
parameters: the mass and width of the Z boson $\MZ$ and $\GZ$, 
the hadronic pole cross-section $\shad=12\pi\Ge\Ghad/(\MZ^2\GZ^2)$,
the ratio of the hadronic to leptonic widths $\Rl=\Ghad/\Gl$ and the
pole forward-backward asymmetry for leptons $\Afbzl$. These are
chosen to be largely uncorrelated experimentally.
A more detailed discussion
of the variables and their definitions can be found in \cite{pbrrev}.
The five parameter formalism assumes {\it lepton universality}. 
If this hypothesis
is not imposed then the results are given separately for $\Rl$ and $\Afbzl$
for each lepton species; a total of nine parameters.

 The above quantities have been accurately determined by the LEP experiments 
using the large statistics gathered at, or close to, the Z peak. 
At the SLC the polarised 
electron beam gives additional information on the couplings from a
measurement of
\begin{equation}
\ALR = \frac{\sigmal - \sigmar} { \sigmal + \sigmar} = \cAe \ \ ,
\end{equation}
 where $\sigmal$($\sigmar$) is the total cross-section for a left (right)
handed polarised incident electron beam. 
This measurement is inclusive to any Z final state. If the forward-backward
asymmetry for a fermion f is also measured for the two polarisation states,
giving the left-right forward-backward asymmetry $\Alrfb$,
then the coupling $\Aferm$ can be directly extracted. Measurements of all these
quantities have been performed by the SLD Collaboration.

 All of these results can be combined and values of the individual couplings
of the leptons and b and c quarks to the Z can be extracted. The SLD results
give directly $\cAe$, $\Ab$ and $\Ac$; thus giving the ratios of the
vector to axial-vector couplings without the use of other data. The
$\tau$ polarisation gives directly values of $\cAe$ and A$_{\tau}$
The forward-backward asymmetries obtained using unpolarised beams give
the product of $\cAe$ and $\Aferm$, so the extracted values of $\Ab$ and $\Ac$
depend on $\cAe$.

\section{Results on Z boson couplings}

 The data collection at LEP at, and around, the Z-boson mass (LEP 1 phase) 
finished in 1995.
Although most of the data have been analysed, only preliminary results are 
generally available, so should be treated with a degree of caution.
The total luminosity recorded by each LEP experiment is 
about 160~pb$^{-1}$; corresponding to about 5 million Z decays. 
The data used in the fits below are from \cite{ewwg97}.

\subsection{Lineshape and lepton asymmetries}

 The results of the nine and five parameter fits to the individual LEP 
experiments are combined, taking into account common systematic errors.
These are given in Tables~\ref{T:9par} and ~\ref{T:5par} respectively.
The most important of these are the LEP energy errors and the theoretical
uncertainty on the luminosity. Because of the complicated correlations between
the experimental systematic errors between years and between energy points
and also those of the LEP energy\cite{lepener}, the evaluation of 
the components of the
final errors on $\MZ$ and $\GZ$ arising from the LEP energy uncertainty is
not straightforward. However, an estimate of this has been made by the
LEP Electroweak Working Group \cite{ewwg97}, 
giving $\delta \MZ(LEP) = \pm$ 1.7 MeV and $\delta \GZ(LEP) = \pm$ 1.3 MeV.

\begin{table}[htb]
\small
\begin{center}

\begin{tabular}{|l||l|l||rrrrrrrrr|}\hline
quantity& value & error & $\MZ$
& $\GZ$ & $\shad$ & $\Ree$ & $\Rmu$ & $\Rtau$& $\Afbze$& $\Afbzm$& $\Afbzt$ \\
\hline
$\MZ$(GeV)  & 91.1867  & 0.0021 &  1.000 &  0.000 & -0.040 & 0.002  &  -0.010
& -0.006 & 0.016 & 0.045 & 0.038 \\
$\GZ$(GeV)  &  2.4939  & 0.0024 &       &   1.000 & -0.184 & -0.007 &  0.003
& 0.003 & 0.009 & 0.000 & 0.003 \\
$\shad$(nb) & 41.491   & 0.058  &       &        &  1.000 & 0.058  &  0.094
& 0.070 & 0.006 & 0.002 & 0.005 \\
$\Ree$       & 20.783   & 0.052  &      &        &       &  1.000 &  0.098
& 0.073 & -0.442 & 0.007 & 0.012 \\
$\Rmu$       & 20.789   & 0.034  &      &        &       &    & 1.000 &  0.105
& 0.001 & 0.010 & -0.001 \\
$\Rtau$        & 20.764   & 0.045  &       &     &      &   &  & 1.000 &  0.002
& 0.000 & 0.020\\
$\Afbze$     & 0.0153   & 0.0025 &   &   &   &  &  &  & 1.000 & -0.008 &-0.006 \\
$\Afbzm$     & 0.0164   & 0.0013 &  &  &  &  &  &  &  & 1.000& 0.029  \\
$\Afbzt$     & 0.0183   & 0.0017 &  &  &  &  &  &  &  &     & 1.000  \\
 \hline
\end{tabular}
\end{center}
 \caption{Results and correlation matrix of the 9 parameter fit to the
LEP data. The $\chit$/df of the average is 28/27, a probability
of 41\%.}\label{T:9par}
\end{table}

\begin{table}[htb]
\vspace*{0.5cm}
\small
\begin{center}

\begin{tabular}{|l||l|l||rrrrr|}\hline
quantity& value & error & $\MZ$
& $\GZ$ & $\shad$ & $\Rl$ &$\Afbzl$  \\
\hline
$\MZ$(GeV)  & 91.1867  & 0.0021 & 1.000 &  0.000 & -0.040 & -0.010 &  0.062  \\
$\GZ$(GeV)  &  2.4939  & 0.0024 &       &  1.000 & -0.184 &  0.002 &  0.004  \\
$\shad$(nb) & 41.491   & 0.058  &       &        &  1.000 & 0.123  &  0.006  \\
$\Rl$       & 20.765   & 0.026  &       &        &        &  1.000 & -0.072  \\
$\Afbzl$    & 0.01683  & 0.00096&       &        &        &        &  1.000  \\
\hline
\end{tabular}
\end{center}
\caption{Results and correlation matrix of the 5 parameter fit to the
LEP data. The $\chit$/df of the average is 31/31, a probability
of 47\%.}
\label{T:5par}
\end{table}

\subsection{$\tau$ Polarisation}

The main improvements in the data in the last year are from the 
very precise results 
(still preliminary) from the ALEPH Collaboration on their full LEP 1 data set. 
The statistical and systematic errors on A$_{\tau}$ are comparable in 
magnitude,  but for the asymmetry measurement  A$_{e}$ the statistical 
errors dominate.
The results of the averaged values for $\cAt$ and $\cAe$ are \cite{tau98}
\begin{eqnarray}
  \cAt & = & 0.1431 \pm 0.0045 \\
   \cAe & = & 0.1479 \pm 0.0051 \,,
\end{eqnarray}
are compatible, in agreement with lepton universality. Assuming
$\mathrm{e}-\tau$ universality, the values for $\cAt$ and $\cAe$ can
be combined. This combination is performed neglecting any possible
common systematic error between $\cAt$ and $\cAe$ within a given
experiment, as these errors are also estimated to be small.  The
combined result of $\cAt$ and $\cAe$ is:
\begin{eqnarray}
  \cAl & = & 0.1452 \pm 0.0034 \,.
\end{eqnarray}

\subsection{Left-Right Asymmetry $\ALR$}

 The high values of longitudinal polarisation (P$_{e}\simeq$ 80 \%)
achieved at the SLC have allowed the SLD experiment to make an extremely
precise, but still preliminary, measurement of
\begin{equation}
\ALR = \frac{\sigmal - \sigmar} { \sigmal + \sigmar} = \cAe \ \ ,
\end{equation}
 where $\sigmal$($\sigmar$) is the total cross-section for a left (right)
handed polarised incident electron beam. 
About 18 pb$^{-1}$ of data have
been collected and analysed,up to and including 1998,
giving ~\cite{SLDALR}
\begin{equation}
\Aelec = 0.15042 \pm 0.00228, \hspace*{1.0cm} \swsqeffl = 0.23109 \pm 0.00029.
\end{equation}
This value is compatible with the less precise value of $\Aelec$
from $\tau$-polarisation
at the 0.5$\sigma$ level, and at the 1.3$\sigma$ level compared to the
value of $\Aell$ from $\tau$-polarisation if lepton-universality is
assumed. The $\ALR$ result is also compatible with the
 value $\Aelec = 0.1498 \pm 0.0043$ from $\Afbpol$ (assuming lepton
universality) at the 0.1$\sigma$ level.

\subsection{Heavy Flavours}

 Extracting electroweak results from heavy flavour data is a rather
involved procedure. This is because knowledge is required of the
various c-quark and b-quark hadron lifetimes, multiplicities,
branching ratios and fragmentation properties. 
The quantities of interest are $\Rbbz = \Gb/\Ghad$, $\Rccz = \Gc/\Ghad$,
and the pole forward-backward asymmetries for b and c 
quarks $\Afbzb$ and $\Afbzc$. At the SLD the left-right forward-backward
asymmetry $\Alrfb$ is also measured for $\bb$ and $\cc$ final states, and these
give direct measurements of $\cAb$ and $\cAc$ respectively.

 For $\Rbbz$ and $\Rccz$ the most reliable and accurate methods exploit
double tags. The number of single and double tags for a b (or c) quark is
found and this can be used to determine both the tagging efficiency
and $\Rbbz$ (or $\Rccz$). Experimentally these measurements required
a b(c) quark tag of high purity and efficiency. Since the main backgrounds 
for b-quark tagged samples are mainly from c-quarks, and vice-versa, it is clear
that the background contamination in these samples must be reliably known.
The light-quark background must still be taken from
Monte Carlo simulations and the hemisphere correlations in the double-tags
must be carefully evaluated. For the forward-backward asymmetries of heavy
quarks use is made of lepton tags, D-meson tags and lifetime tags 
plus jet-charge measurements (see \cite{ewwg97} for details).
In these measurements, particularly for $\Rbbz$, the experimental error
has a large component from systematic effects. 

 The main changes in the data in the last year are new results on $\Rbbz$ from 
DELPHI, OPAL and SLD; on $\Rccz$ from DELPHI, OPAL and SLD,
on $\Afbzb$ from ALEPH and DELPHI and on $\Afbzc$ from
ALEPH, DELPHI and OPAL. In addition, the SLD results on $\cAb$ and $\cAc$ have 
been updated. 

\begin{table}[htb]
\small
\begin{center}
\begin{tabular}{l|l|l||rrrrrr}\hline
quantity& value & error & $\Rbbz$
& $\Rccz$ & $\Afbzb$ & $\Afbzc$ &$\cAb$ & $\cAc$ \\
\hline
$\Rbbz$   &0.21656& 0.00074& 1.000&--0.17& -0.06&  0.02&--0.02&  0.02\\
$\Rccz$   &0.1735 & 0.0044 & & 1.000&  0.05&--0.04&  0.01&--0.04\\
$\Afbzb$  &0.0990 & 0.0021 &  &  & 1.000&  0.13&  0.03&  0.02\\
$\Afbzc$  &0.0709 & 0.0044 &  & &  & 1.000& --0.01&  0.07\\
$\cAb$    &0.867  & 0.035  & &  &  & & 1.000&  0.04\\
$\cAc$    &0.647  & 0.040  &  & & &  &  & 1.000\\
 \hline
\end{tabular}
\end{center}
\caption{
Results of fits to the LEP and SLD heavy flavour data, plus
the correlation matrix.
The $\chit$/df of the average is 44.1/(88-13), a probability
of more than 99$\%$.}
\label{T:hflav}
\end{table}

 The combination of results has been carried out by a LEP/SLD working
group using the procedure described in~\cite{lephf}.
Each experiment provides, for each measurement, a complete breakdown of
the systematic errors, adjusted if necessary to agreed meanings of
these errors. Direct measurements of $\cAb$ and $\cAc$ by SLD,
obtained by measuring $\Afbzb$ and $\Afbzc$ with a polarised beam, are also
included.
A multi-parameter fit is then performed to get the best overall
values of $\Rbbz$,$\Rccz$,$\Afbzb$, $\Afbzc$, $\cAb$ and $\cAc$,
plus their covariance matrix. The results of a fit to both the LEP and SLD
data are given in Table~\ref{T:hflav}. The effective
mixing parameter $\chibar$ and the leptonic branching
ratios b$\rightarrow\ell$ and b$\rightarrow$c$\rightarrow\ell$ are
also included in the fit. It can be seen from Table~\ref{T:hflav}
that the data are very compatible; indeed, there is an indication 
from the overall $\chit$ that
some errors maybe overestimated. The individual 
measurements of $\Afbzb$ are also very compatible\cite{Halley}.

 The results for both $\Rbbz$ and $\Rccz$ are now reasonably 
compatible with the SM. This is
in contrast to the situation in 1995~\cite{beijing} when the results had only
a 1\% confidence level of being compatible with the SM. The changes arise from
several sources. For $\Rccz$, the results no longer depend on the assumption
of the energy dependence of the D-meson production rates, since these are now
measured accurately at LEP. The use of double-tag methods for $\Rccz$ has
also improved. Better tags have also been developed for b-quarks, incorporating
invariant mass and other information. Also the understanding of the vertex
detectors has improved such that efficient, but extremely pure, b-quark tags
are now possible.

An alternative approach of the discussion of the heavy flavour data is
given below.

\section{Extraction of fermion couplings}

 A simultaneous fit is made to the data discussed above in order to
extract both the lepton and heavy quark vector and axial-vector 
couplings\footnote{See also ~\cite{beijing},
\cite{pbrrev},\cite{pbr9804} and ~\cite{field}.}.
The measurements used are the 9 parameter lineshape results (which
reduces to 5 parameters if lepton universality is imposed),
the $\tau$-polarisation results for $\cAe$ and $\cAt$, the SLD measurement
of $\cAe$ and the 6 parameter heavy flavour results.

The main information content in these measurements is 
from $\Rbbz = \Gb/\Ghad$ (which, using $\Ghad$ from the
lineshape, gives $\Vb^{2}$+ $\Abcoup^{2}$), $\Rccz$
($\Vc^{2}$ +  $\Accoup^{2}$), $\Aelec$ from LEP/SLD ($\Ve/\Ae$),
$\Afbzb$ ($\Vb/\Abcoup$, $\Ve/\Ae$), $\cAb$ ($\Vb/\Abcoup$),
 $\Afbzc$ ($\Vc/\Accoup$, $\Ve/\Ae$) and $\cAc$ ($\Vc/\Accoup$). The constraint
$\alphasmz$ = 0.119 $\pm$ 0.003 is imposed (although the results are rather
insensitive to this as discussed below).

 If lepton universality is not assumed then information on the lepton couplings 
comes from the direct measurements of $\cAe$ and $\cAt$, as well as from
the lepton forward-backward asymmetries and lepton partial widths and also
from the heavy quark forward-backward asymmetries. These
are contained in the 9 parameter lineshape results and the 6 parameter
heavy flavour results respectively. Through correlations the other parameters 
also enter in these fits. 
The overall $\chit$ of the fit is 2.3 for 5 df, giving a probability of 81\%.
If lepton universality is  assumed then the $\chit$ of the fit becomes
5.7 for 4 df, giving a probability of 22\%.

\subsection{Lepton couplings}

\begin{figure}[htpb!] 
\centerline{\epsfig{file=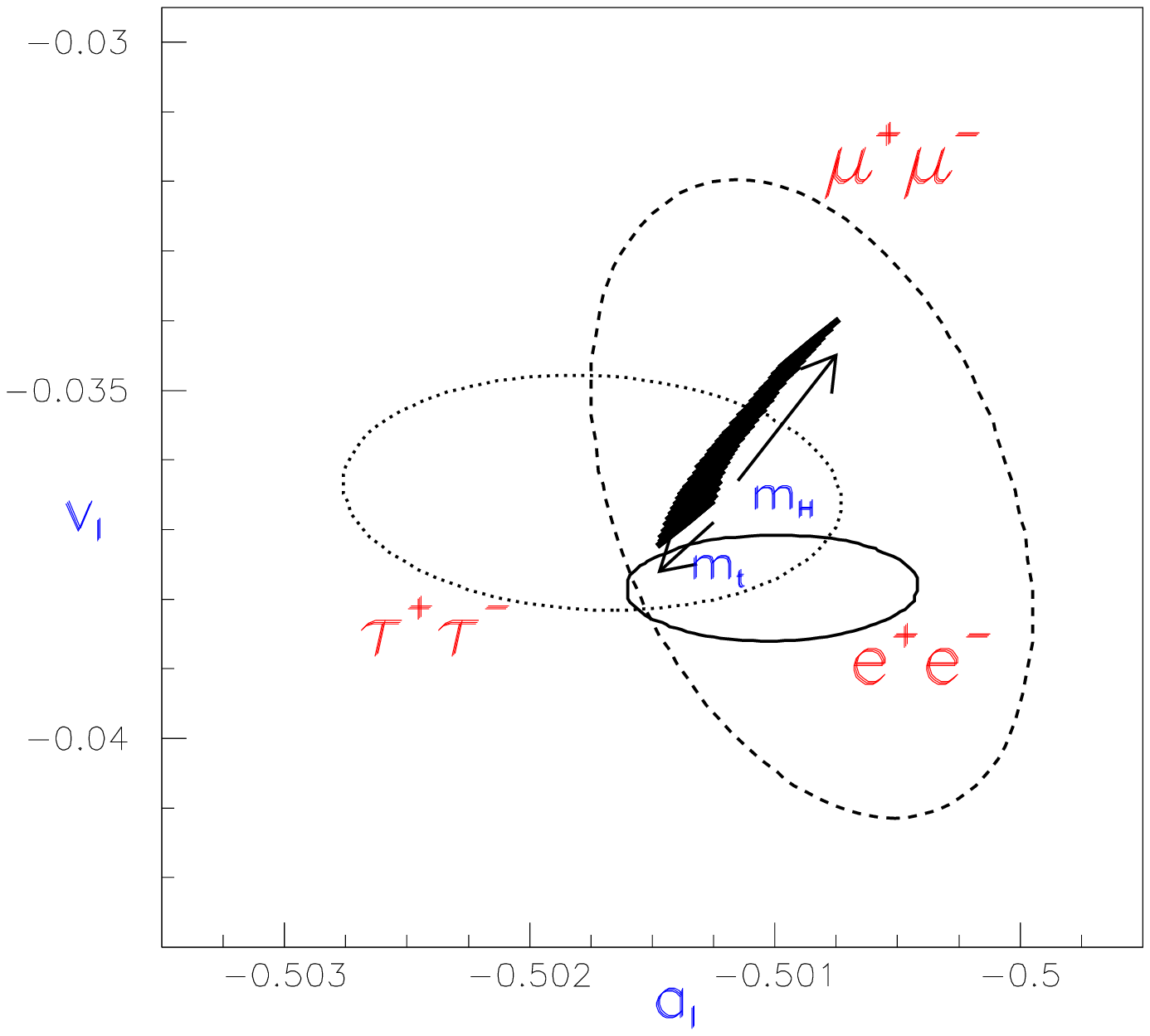,height=3.5in,width=3.5in}}
\vspace{10pt}
\caption{
 Contours of 70\% probability in the $\Vlll$-$\Alll$ plane from
  LEP and SLD measurements. The solid region
  corresponds to the Standard Model prediction for $\tI$ GeV
  and  $\HI$ GeV. The arrows point in
  the direction of increasing values of $\Mt$ and $\MH$.  }
\label{figleptuniv}
\end{figure}

 {\it Lepton universality} is a hypothesis of the SM and it is clearly important
to test it as precisely as possible.  The results of the fit for the
individual lepton couplings are
shown in Fig.\ref{figleptuniv}, together with the 70\% confidence level
contours. 
The signs are plotted taking $\Ae <$ 0. Using this convention (this is
justified from $\nu$-electron scattering results \cite{Winter95}), the signs 
of all couplings are uniquely determined from LEP data alone.
The data are compatible with the hypothesis of {\it lepton universality}
and with the SM expectations.
The results test this hypothesis at the level of 0.1\% for $\Alll$, but only at 
the 5-10\% level for $\Vlll$. 

\begin{table}[htb]
\small
\begin{center}
\begin{tabular}{|l|c|c|}\hline
   & LEP  & LEP+SLD \\
\hline
$\Vlll$   & -0.03719 $\pm$ 0.00061  &  -0.03756 $\pm$ 0.00042  \\
$\Alll$   & -0.50107 $\pm$ 0.00030  &  -0.50105 $\pm$ 0.00030  \\
$\Vnu$   & +0.50127 $\pm$ 0.00095  &  +0.50128 $\pm$ 0.00095  \\ 
\hline
\end{tabular}
\end{center}
\caption{Lepton vector and axial-vector couplings assuming lepton 
universality.}
\label{T:leptcoup}
\end{table}

\begin{figure}[htbp!] 
\centerline{\epsfig{file=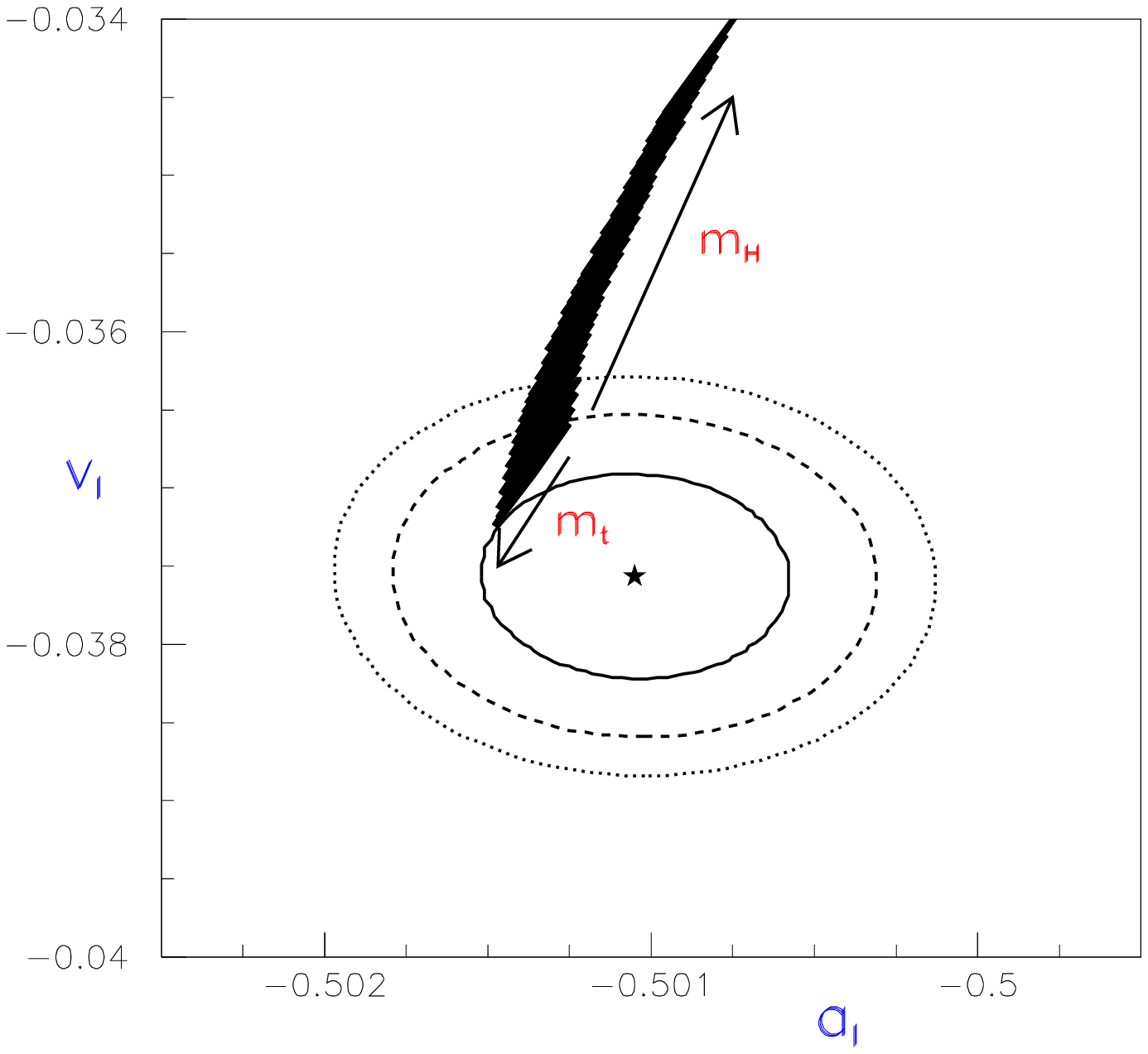,height=3.6in,width=3.6in}}
\vspace{10pt}
\caption{
 Contours of 70\%, 97\% and 99\%  probability in the $\Vlll$-$\Alll$ plane from
  LEP and SLD measurements. The solid region
  corresponds to the Standard Model prediction for $\tI$ GeV
  and  $\HI$ GeV. The arrows point in
  the direction of increasing values of $\Mt$ and $\MH$.  }
\label{figvae}
\end{figure}

\begin{figure}[htb!] 
\centerline{\epsfig{file=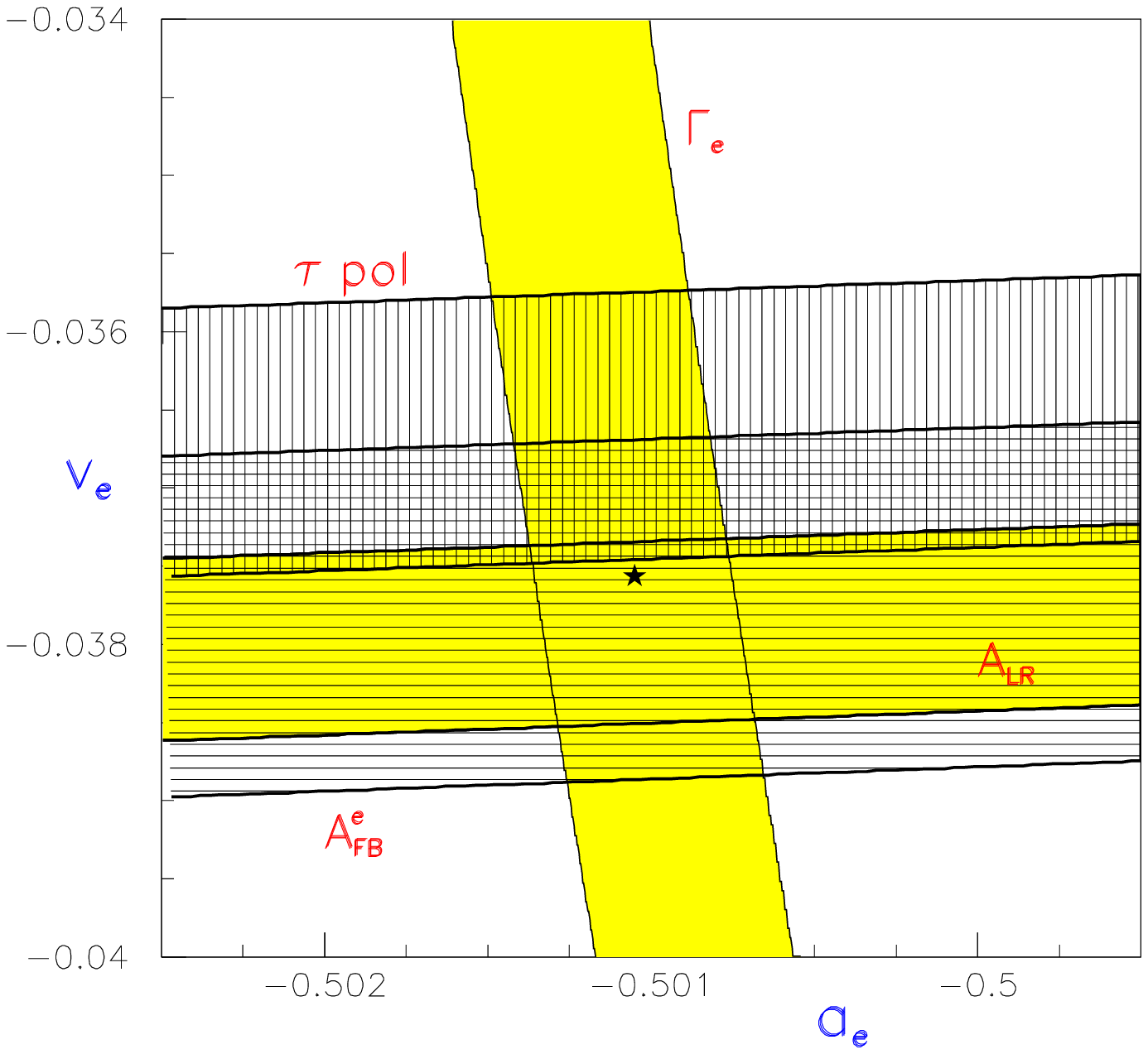,height=3.8in,width=3.8in}}
\vspace{10pt}
\caption{
Constraints on $\Vlll$ and $\Alll$ from individual measurements. }
\label{fig-indive}
\end{figure}

 If the hypothesis is assumed then the
computed couplings are given in Table~\ref{T:leptcoup} and shown 
in Fig.\ref{figvae}. The constraints from the individual measurements are
shown in  Fig.\ref{fig-indive}. It can be seen that $\Alll$ is essentially
determined by $\Gl$, whereas $\Vlll$ is determined, in order of accuracy,
by the measurements of $\ALR$, the $\tau$-polarisation and $\Afbpol$.
In the context of the SM these measurements, in particular $\ALR$, favour 
a rather light Higgs boson mass.

\subsection{Heavy quark couplings}

\begin{figure}[htb!] 
\centerline{\epsfig{file=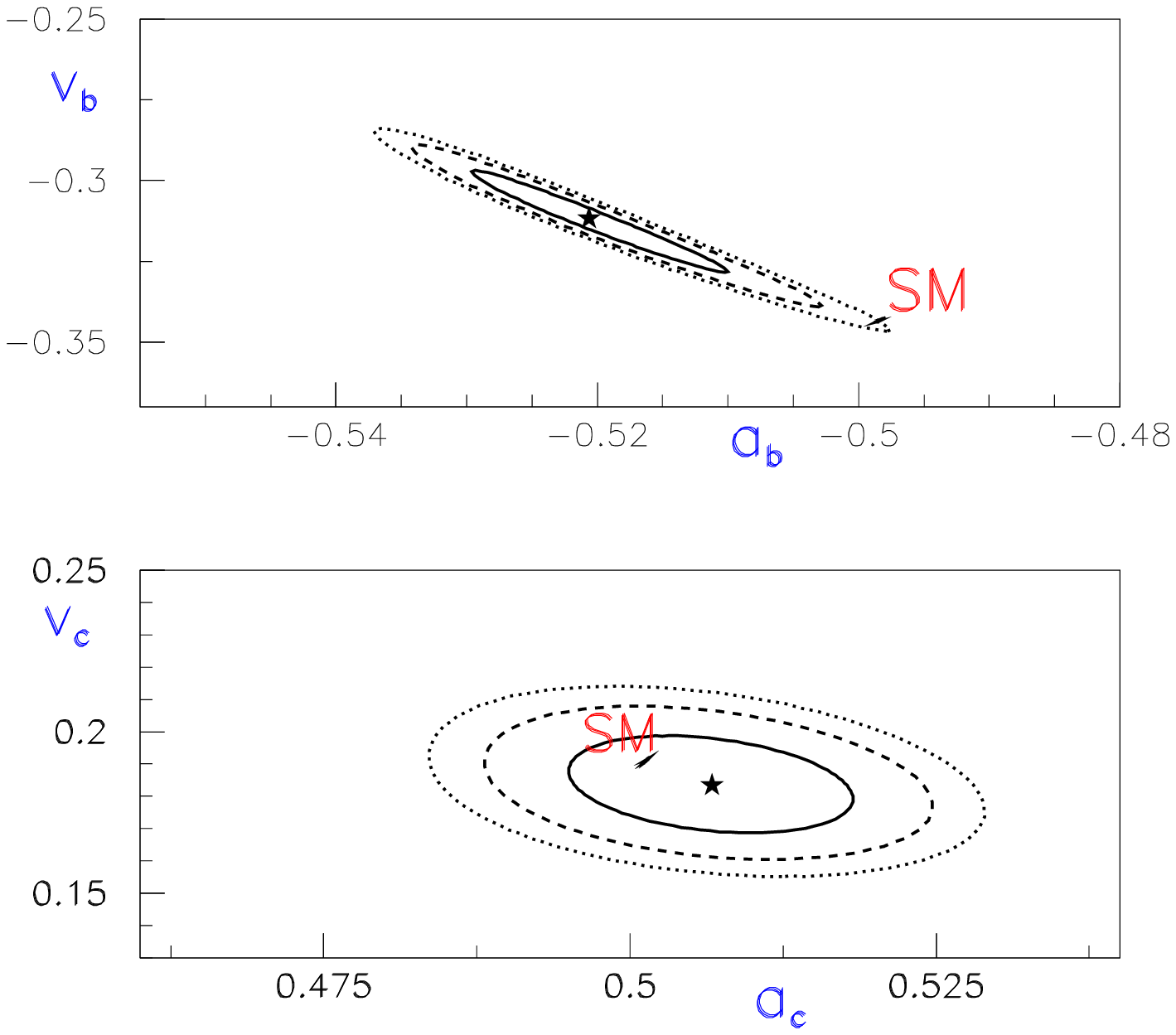,height=4.0in,width=4.0in}}
\vspace{10pt}
\caption{
Results of a fit to the b and c-quark vector and axial-vector couplings.
The contours are for the 70, 95 and 99\% confidence limits.  }
\label{fig-coupvab}
\end{figure}

\begin{figure}[htb!] 
\centerline{\epsfig{file=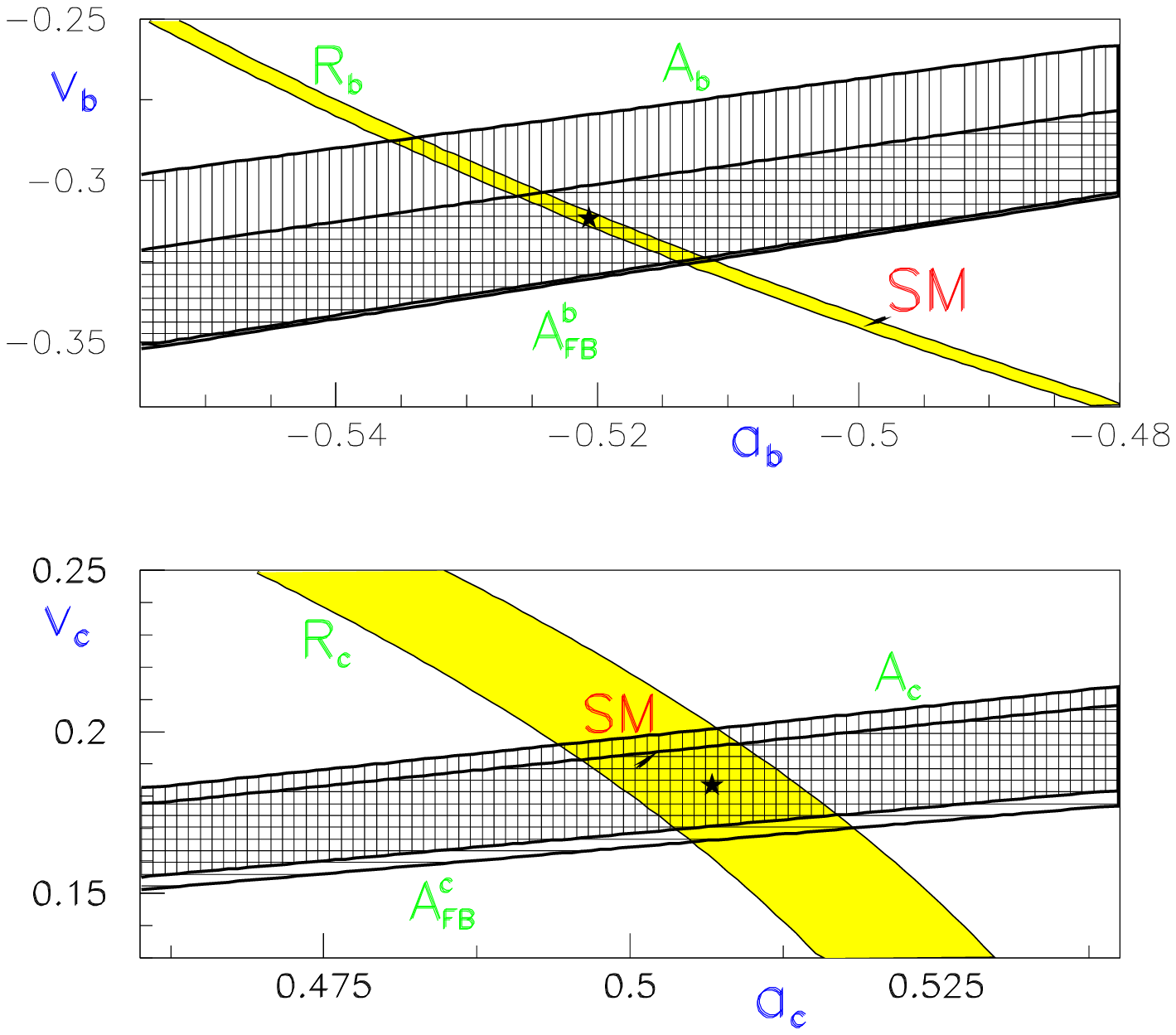,height=4.0in,width=4.0in}}
\vspace{10pt}
\caption{
Constraints on the b and c-quark vector and axial-vector couplings,
from individual measurements.
 }
\label{fig-indiv}
\end{figure}

The results of the fit for $\Vb$ and $\Abcoup$ are given 
in  Table ~\ref{T:hfva} and Fig.~\ref{fig-coupvab}. Also shown are
the SM predictions corresponding to $\Mt=173.8\pm5.0$~GeV and $\HI$.
The corresponding results for $\Vc$ and $\Accoup$ are also shown.
Note that there
is a very strong anti-correlation between $\Vb$ and $\Abcoup$.
 The constraints from the individual measurements are shown in 
Fig.~\ref{fig-indiv}.
 
\begin{table}[hbp]
\small
\begin{center}
\begin{tabular}{|c|c|rrrr|}\hline
parameter & fitted value  & $\Vb$ &$\Abcoup$ &$\Vc$ &$\Accoup$  \\
\hline
$\Vb$     & $-0.3118\pm0.0101$   & 1.00 & -0.98 & -0.15 & 0.04  \\
$\Abcoup$ & $-0.5206\pm0.0063$ &      & 1.00  & 0.15  & -0.01 \\
$\Vc$     & $0.183\pm0.010$    &      &       & 1.00  & -0.29   \\
$\Accoup$ & $0.5067\pm0.0075$  &      &       &       & 1.00 \\
\hline
\end{tabular}
\end{center}
\caption{Results, plus correlation matrix, of a fit to the vector and
axial-vector couplings of b and c quarks.}
\label{T:hfva}
\end{table}

\begin{figure}[htb!] 
\centerline{\epsfig{file=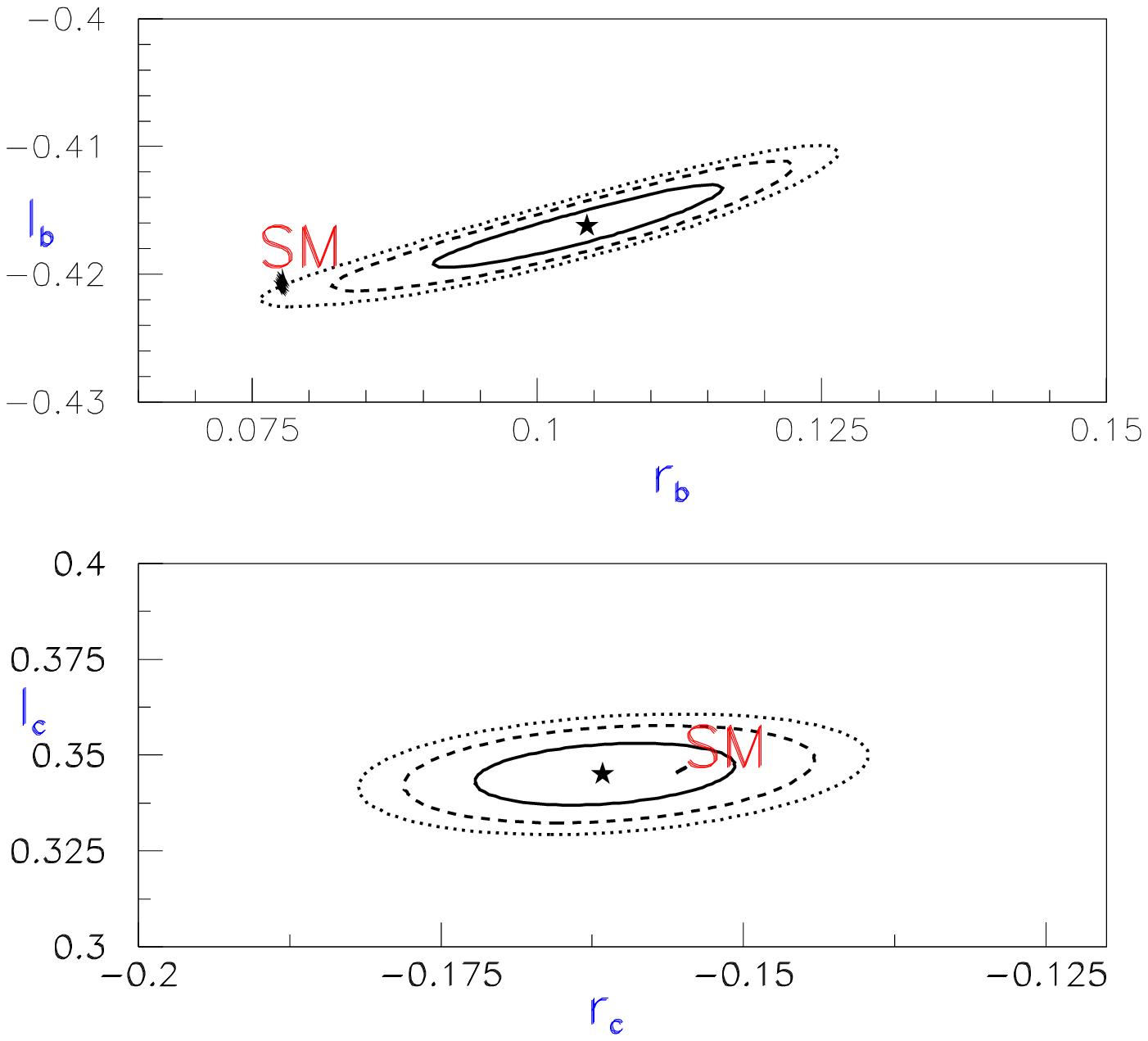,height=4.0in,width=4.0in}}
\vspace{10pt}
\caption{
Results of a fit to the b-quark left and right-handed couplings.
The contours are for the 70, 95 and 99\% confidence limits.  }
\label{fig-couplrb}
\end{figure}

 The b(or c)-quark couplings can also be expressed in terms of the
left-handed  $\ell_{b}$ = ($\Vb$ + $\Abcoup$)/2 and
right-handed r$_{b}$ = ($\Vb$ - $\Abcoup$)/2 couplings.
The results are shown in Fig.~\ref{fig-couplrb}. 
It can be seen that, whereas the c-quark couplings are reasonably 
compatible with the
SM, those for the b-quark, in particular the right-handed coupling, are
are in poor agreement with the SM expectations. 
The fitted values of $\Vb$ and $\Abcoup$ (or  $\ell_{b}$ and  r$_{b}$)
give a value of $\Rbbz$ greater than the SM value,
and a value of $\cAb$ (or $\Afbzb$) less than the SM value. In that
sense the b-quark data are mutually consistent with the observed deviations
from the ~SM.  

\subsection{Discussion of results}

 The results for the Z-lepton couplings from the different methods
are compatible both with each other\footnote{The results now are more
consistent than in previous years; see e.g. \cite{beijing},\cite{pbrrev}.} and 
with the SM, provided the Higgs Boson is relatively light. 
The c-quark couplings are also compatible with the SM. However, as can
be seen from Fig.~\ref{fig-coupvab}, the b-quark couplings appear to be
only marginally compatible with the SM. The point in the SM band which is 
closest to the fitted data values corresponds to $\Mt$ = 168.8 GeV and
$\MH$ = 90 GeV. The $\chit$ probability for compatibility to this point
is 1.3\%. 

 It is worthwhile therefore exploring further this possible discrepancy.
In the fits the assumed value of $\alphasmz$ was taken to 
be 0.119 $\pm$ 0.003. If a central value of 0.116 is used then the leptonic
couplings are unchanged and the shifts in the b- and c-quark couplings 
are 0.0003 or less. Hence the results are not very sensitive to $\alphasmz$.
This is to be expected since the ratios $\Rbbz$ and $\Rccz$ are, by
construction, rather insensitive to $\alphasmz$.

 The results from the SLD Collaboration on $\ALR$, $\Ab$ and
$\Ac$ \cite{SLDALR} require a precise determination of 
the degree of polarisation of the
electron beam. It can be noted that the values of $\Aelec$ (from  $\ALR$),
$\Ab$ (from $\Alrfbb$) are above 
and below the SM predictions respectively. Since, in both cases, what is 
measured is 
proportional to the product of the polarisation and the required parameter,
the measurements cannot both be reconciled with the SM simply by a change in the
value of the electron polarisation. It is worth stressing that the uncertainty
on $\ALR$ due to the  polarisation is between 0.7\% and 1.1\% for the 
published and preliminary data sets. This is to be compared to the overall
statistical component of the error which is about 1.5\%.

 The results for $\Afbzb$ measure the product of $\Aelec$ and $\Ab$.
Thus the value of $\Abb$ extracted depends critically on that of $\Aelec$.
In the standard fits above the information on $\Aelec$ comes from all of
the data and the fitted value is $\Aelec$ ~= 0.1491 $\pm$ 0.0017. Most of
the information comes from the measurements of $\ALR$, the $\tau$-polarisation
and $\Afbzl$. In the SM the value of $\Aelec$ increases for 
increasing $\Mt$ and decreasing $\MH$. However, as $\Mt$ is now well
constrained, the main variation is from  $\MH$. As can be seen from 
Fig.~\ref{figvae}, the lepton coupling data favour a light Higgs.
Within the ranges $\tI$ and $\HI$  the closest SM value is 0.1477, which
corresponds to $\Mt$ = 178.8 GeV and $\MH$ = 90 GeV. The value of $\Aelec$
which corresponds to the 95\% upper limit on $\MH$ (namely about 300 GeV)
is 0.1460. The values of $\Vb$ and  $\Abcoup$ extracted when these two values
for $\Aelec$ are imposed for the measurement of $\Afbzb$ are given in
Table~\ref{T:vabvalues}. Also given are the $\chit$ probabilities that
the result is compatible with the closest SM value.
It can be seen that the $\chit$ probability increases
as  $\Aelec$ increases to about 5\% for $\Aelec$ = 0.1460. If $\Afbzb$ is
removed from the fit then the  probability increases to 6.6\%.

 Although the c-quark couplings are resonably compatible with the SM
it can be noted that, as for b-quarks, the measured values of $\Afbzc$
and $\Ac$ are both below the SM predictions. However, the errors for
c-quarks are currently larger than for  b-quarks, so the differences
are less significant.

\begin{table}[htb]
\vspace*{0.5cm}
\small
\begin{center}
\begin{tabular}{|c|c|c|c|}\hline
 conditions on $\Afbzb$  & $\Vb$  & $\Abcoup$ & $\chit$ 
 prob. for SM \\
\hline
 none  & -0.3118 $\pm$ 0.0101 & -0.5206 $\pm$ 0.0063 & 1.3\% \\
 $\Aelec$ = 0.1477   & -0.3154 $\pm$ 0.0094 & -0.5185 $\pm$ 0.0059 & 1.7\% \\
 $\Aelec$ = 0.1460   & -0.3199 $\pm$ 0.0097 & -0.5157 $\pm$ 0.0062 & 4.8\% \\
 remove              & -0.3147 $\pm$ 0.0119 & -0.5189 $\pm$ 0.0075 & 6.6\% \\
\hline
\end{tabular}
\end{center}
\caption{ Values of $\Vb$ and $\Abcoup$ for different assumptions 
about  the use of $\Afbzb$.
The SM values used correspond to $\Mt$ = 168.8 GeV and $\MH$ = 90 GeV.}
\label{T:vabvalues}
\end{table}

\section{Electroweak fits for the Higgs boson mass}

 The precision data discussed above, together with additional data,
are sufficiently accurate to constrain the mass of the  Higgs boson 
in electroweak fits.
The additional measurements used are $\MW$= 80.390 $\pm$ 0.064 GeV, 
$\swsqa$ = 0.2254 $\pm$ 0.0021 from deep inelastic neutrino-nucleon experiments
and $\swsqeffl$ = 0.2321 $\pm$ 0.0010 from the flavour averaged 
forward-backward asymmetry in Z hadronic events~$\avQfb$\cite{ewwg97}. 
The value of  $\MW$  is the current average of the Fermilab and CERN values.
The value of $\swsqa$ is from the NuTeV\cite{NuTeV} and CCFR\cite{CCFR} 
experiments, and the dependence of the result on $\Mt$ and $\MH$\cite{NuTeV}
is taken into account. The top quark mass~\cite{topmass} 
$\Mt=173.8\pm5.0$~GeV is used as a constraint in the fits.

The parameters used in these
fits are $\MZ$, $\Mt$, $\alphasmz$ and $\alphamz$ as well as the Higgs
mass $\MH$. An external constraint  1/$\alpha^{5}(\MZ) = 128.878  \pm 0.090$
~\cite{Eid95} is used in the fits discussed below.

\begin{table}[htb]
\vspace*{0.5cm}
\small
\begin{center}
\begin{tabular}{|l|c|c|c|c|}\hline
  data used & $\Mt$ GeV & $\MH$ GeV & 95\% limit on $\MH$ GeV & 
$\chit$ prob. of fit  \\
\hline
all data   & 171.1$^{+4.9}_{-4.8}$  &  77$^{+85}_{-47}$ & 246 & 35\% \\
without $\ALR$   & 172.5$^{+4.9}_{-4.8}$  &  150$^{+133}_{-79}$ & 410 & 63\%  \\
without $\Afbzb$   & 169.7$^{+4.8}_{-4.2}$  &  36$^{+66}_{-22}$ & 172 & 56\%  \\
without $\ALR$ and $\Afbzb$ & 171.7$^{+5.0}_{-4.9}$  &  105$^{+124}_{-70}$ 
& 348 & 65\%  \\
scale $\ALR$ and $\Afbzb$ & 171.5$^{+4.9}_{-4.9}$  &  92$^{+106}_{-61}$ 
& 303 & 70\%  \\
\hline
\end{tabular}
\end{center}
\caption{ Results of electroweak fits to $\Mt$ and $\MH$ for different
sets of data. The fourth column gives the 95\% one-sided confidence level
upper limit on $\MH$. This upper limit does not include the uncertainty
in the theory. A fitted value of $\alphasmz$ = 0.119 $\pm$ 0.003 is also
obtained in these fits.}
\label{T:higgs}
\end{table}

 The results of the fits to all electroweak data give a central value
for $\MH$ of 77~GeV, and a one-sided 95\% c.l. upper limit of 246 GeV;
see Table ~\ref{T:higgs}.\footnote{The fits have been made 
using ZFITTER version 5.12.} The lower
limit from direct searches of about 90 GeV~\cite{Treille} is not used
in the limits here. 
The upper limit does not take into account the theoretical uncertainty 
due to missing higher order terms. Including an estimate of these (as discussed
in \cite{ewwg97}) increases
this limit increases to 262 GeV; that is, an increase of about 16 GeV.
It is of great importance, particularly in the consideration of the 
construction of new accelerators,
to understand if these values are reliable. One can adopt (at least) two 
approaches to these fits:

\begin{itemize}
\item[1)] The overall $\chit$ of 14/13 d.f. (prob. = 35\%) for the fit
to all data is
reasonably good. The distribution of the pulls\footnote{The pull is defined
as the difference between the measured and fitted values, divided by the
error on the quantity.}
has a mean value
of -~0.1 ~$\pm$ 0.2 and an rms of 0.9, and so is compatible with
the expected Gaussian distribution. The two measurements with the largest
$\chit$'s ($\Afbzb$ and $\ALR$) are just the expected ``tails'' of the
distribution. However, these are the two most sensitive measurements 
to $\MH$.
 
\item[2)] The quantities which are most sensitive to $\MH$
are, in order of current  sensitivity, $\ALR$, $\Afbzb$,$\GZ$, 
P$_{\tau}$, $\MW$ and $\Afbzl$. 
These 6 quantities contribute 7.5 to the $\chit$. The individual values
of the pulls for these 6 quantities are -1.7, -1.8, -0.8, -0.4, 0.3 and
0.7 respectively.
The central value for $\MH$ is sensitive to which data are included.
For example, if a fit is performed without the inclusion of $\ALR$, the most
sensitive quantity to $\MH$,
then, as shown in Table~\ref{T:higgs}, the one-sided 95\% c.l. upper limit 
increases to 410 GeV, plus the theory error.
However if $\Afbzb$ (alone) is excluded
from the fit, then the 95\% c.l. upper limit on $\MH$ becomes 172 GeV.
If both $\ALR$ and $\Afbzb$ are excluded, then the 95\% c.l. upper limit
on $\MH$ becomes 348 GeV.
 Although the 6 most sensitive quantities to $\MH$ have a reasonable
total contribution to $\chit$, the two most sensitive quantities, 
$\ALR$ and $\Afbzb$ have a $\chit$ contribution of 6.0. If the errors on these
two quantities are scaled according to the Particle Data Book recipe, then
the 95\% c.l. upper limit on $\MH$ becomes 303 GeV.

\end{itemize}

 Of particular interest is the extent to which the data indicate that the
Higgs is light. As can be seen from the fits above, the quantity most 
responsible for driving the limit higher is $\Afbzb$. 
However, there is no good reason the reject either the
$\Afbzb$ or $\ALR$ measurements at the present time. The value of $\Ab$, which
contributes to the b-quark couplings not being very compatible with the SM,
has little influence on these fits; apart from increasing the $\chit$. For
example, if the value of $\Ab$ is set to the SM value instead of 
the experimentally measured value, then the Higgs mass changes 
by less than 1 GeV.

 In summary, the best estimate is that the Higgs is relatively light.
However, the data are not fully compatible, so some caution in intrepreting
the data is necessary.

\section{Summary and Conclusions}
 
 The vector and axial-vector couplings of both the leptons and heavy quarks
have been extracted from the most recent electroweak data. The lepton
couplings support the hypothesis of {\it lepton universality}.

 The c-quark couplings are compatible with SM expectations. However, those
of the b-quark agree with the SM at only the 1\% level. The main discrepancy
is for the right-handed coupling of the b-quark. If real, this would not be 
easily interpreted in terms of the usual extension to the SM.
However, it should be noted that the data used are mostly still preliminary
and that the completed final analyses of all the LEP data are eagerly
awaited. Possible future running of the SLC would clearly be of great benefit
in resolving these questions.

 Although the present results on the b-quark couplings are clearly 
interesting, they do
not as yet provide compelling evidence for physics beyond the SM.

\vspace*{2.0cm}

{\bf Acknowlegements}

 I would like to thank my colleagues in the LEP Electroweak Working Group,
in particular M. Grunewald and G. Quast, for valuable discussions.


\end{document}